# Extending the class of solvable potentials
## II. Screened Coulomb potential with a barrier

A. D. Alhaidari[a,b,c,*]

[a]*Saudi Center for Theoretical Physics, Dhahran, Saudi Arabia*
[b]*KTeCS, P.O. Box 32741, Jeddah 21438, Saudi Arabia*
[c]*Physics Department, King Fahd University of Petroleum & Minerals, Dhahran 31261, Saudi Arabia*

This is the second article in a series where we succeed in enlarging the class of solvable problems in one and three dimensions. We do that by working in a complete square integrable basis that carries a tridiagonal matrix representation of the wave operator. Consequently, the wave equation becomes equivalent to a three-term recursion relation for the expansion coefficients of the wavefunction in the basis. Finding solutions of the recursion relation is equivalent to solving the original problem. This method gives a larger class of solvable potentials. The usual diagonal representation constraint results in a reduction to the conventional class of solvable potentials. However, the tridiagonal requirement allows only very few and special potentials to be added to the solvability class. In the present work, we obtain S-wave solutions for a three-parameter 1/r singular but short-range potential with a non-orbital barrier and study its energy spectrum. We argue that it could be used as a more appropriate model for the screened Coulomb interaction of an electron with extended molecules. We give also its resonance structure for non-zero angular momentum. Additionally, we plot the phase shift for an electron scattering off a molecule modeled by a set of values of the potential parameters.

## I. INTRODUCTION

Very few problems in quantum mechanics (relativistic and nonrelativistic) are exactly solvable. Despite the limited number of these problems, there are many advantages to obtaining exact solutions (energy spectrum and eigenfunctions) of their associated wave equations using as many methods as possible. One such advantage is that the analysis of such solutions makes the conceptual understanding of quantum physics straightforward and sometimes intuitive. Moreover, these solutions are valuable means for checking and improving models and numerical methods introduced for solving complicated physical systems. In fact, in some limiting cases or for some special circumstances they may constitute analytic models of realistic problems or approximations thereof. Additionally, potentials associated with these solvable problems could be used in the unperturbed part of more realistic Hamiltonians. Consequently, all attempts at enlarging the class of potentials for which an exact solution is obtainable are important and very fruitful.

Aside from the three well-known classes of solvability (exact, conditionally-exact [1], and quasi-exact [2]), we define here the notion of *exact solvability* to be the ability to write the solution of the wave equation in a closed form as a convergent series in terms of quantities that are well-defined to all orders. Moreover, all physical quantities in the

---

problem (e.g., the energy spectrum, phase shift, wavefunction, resonances, etc.) are obtainable to any desired accuracy limited only by the computing machine precision; no physical approximations are involved. A subclass of this notion of *exact solvability* is the *analytic solvability* in which all objects in the resulting expression for any physical quantity are mathematically well-defined in terms of the independent variables (e.g., position, time, angular momentum, energy, potential parameters, etc.). We also reintroduce the concept of a *parameter spectrum* where an exact solution of the problem is obtained at a single energy but for a set (finite or infinite) of values of the potential parameters (the parameter spectrum) [3].

In a recent publication (henceforth referred to as paper I) [4], we used the tools of our "tridiagonal physics" program [5], which was inspired by the J-matrix method [6], to obtain an exact $L^2$ series solution for the infinite potential well with a sinusoidal bottom. No analytic expression was obtained for the energy spectrum formula. However, the accuracy of the calculated values of the energy spectrum is limited only by the computing machine precision; no physical approximations were ever involved. The wavefunction was written as an infinite convergent series whose terms are the product of the trigonometric function in configuration space and the *dipole polynomial* in the energy [7]. In this work, we employ the same technique in paper I to solve a highly significant problem in three dimensions with the following three-parameter central potential

$$V(r) = V_0 \frac{e^{-\lambda r} - \gamma}{e^{\lambda r} - 1}, \qquad (1.1)$$

where $V_0$ is the potential strength and the range parameter $\lambda$ is positive with an inverse length units. The dimensionless parameter $\gamma$ is in the open range $0 < \gamma < 1$. This is a short range potential with $1/r$ singularity at the origin. Figure 1 is a graphical representation of this potential function for $V_0 > 0$ (Fig. 1a) and $V_0 < 0$ (Fig. 1b). The potential trace crosses the radial axis at $r_0 = -\ln(\gamma)/\lambda$ then reaches a local extremum value of $V(r_1)$ $= -V_0 \left(1 - \sqrt{1-\gamma}\right)^2$ at $r_1 = -\frac{1}{\lambda} \ln\left(1 - \sqrt{1-\gamma}\right)$. It is interesting to note that at short distances, there is a clear resemblance between this potential (with $V_0 > 0$) and the attractive Coulomb potential for non-zero angular momentum. However, the potential valley in Fig. 1a is not due to the centrifuge kinematics attributed to non-zero angular momentum. Moreover, the long-range behavior is not the same. Thus, in contrast to the long-range Coulomb potential, we expect that the number of bound states for this potential to be finite. In Fig. 1c, we show the effective potential, which is the sum of $V(r)$ for $V_0 < 0$ and the orbital term with non-zero angular momentum. For certain range of values of the potential parameters and angular momentum, we obtain the configuration shown in the figure with local maximum and minimum. Additionally, with $V_0 < 0$ (see, Fig. 1b) this potential exhibits a rather different type of charge screening. Near the origin, the electron experiences a strong attraction to an *effective nucleus* with effective charge $Z_{eff}$ $= 4\pi\epsilon_0 V_0 (1-\gamma)/\lambda e^2$, which is not necessarily an integer but is less than $Z$ (the sum of all proton charges in the nuclei of the molecule). As the electron gets farther away from the origin, the screening due to the electron cloud around the effective nucleus increases and becomes significant until a balance is reached at $r_1$, which is less than the charge radius of the molecule. Beyond that, a local excess/deficiency of negative/positive charges contributed by electrons/nuclei from the outer/inner atoms in the extended molecule repels the scattered electron until it gets far enough from the center of the molecule where the Coulomb interaction diminishes rapidly. As such, this potential could be used as a more appropriate model for the interaction of an electron with extended molecules whose electron cloud is congregated near the center of the molecule (for example, due to a large centered atom, see Fig. 2). A measure of the molecule extension could be given by the parameter $\sigma = 1 - (Z_{eff}/Z)$. The usual Coulomb screening interaction potentials (e.g., the Hulthén [8] and Yukawa [9] potentials) do not have the barrier structure shown in Fig. 1b and they behave close to the origin like $\frac{Z}{r}e^{-\alpha r}$. In fact, the potential (1.1) exhibits such behavior in the limit $\gamma = 0$, in which case $Z = 4\pi\epsilon_0 V_0/\lambda e^2$ and $\alpha = 2\lambda$. Therefore, the parameter $\gamma$ has the physical interpretation as being the measure of molecular extension when the potential (1.1) is taken as a model for the interaction of an electron with extended molecules. This is because $1 - (Z_{eff}/Z) = \gamma$.

Choosing $\gamma < 0$ or $\gamma > 1$ in the potential (1.1) takes it outside the class of problems under current study. However, our approach can still easily handle the bound states solution for these cases provided that $\gamma V_0 > 0$. In fact, these two cases are simpler and carry less complicated structure as opposed to the case where $0 < \gamma < 1$. For all real $\gamma$ the solution space splits into three disconnected subspaces with different physical interpretations. We limit our investigation here to the middle open range $0 < \gamma < 1$ that has a richer structure. For the boundary case where $\gamma = 1$, the potential (1.1) becomes the simple exponential potential, $-V_0 e^{-\lambda r}$. However, $\gamma = 0$ results in a strongly decaying Hulthén potential, $V_0 e^{-2\lambda r}/(1 - e^{-\lambda r})$ (i.e., heavily screened Coulomb) as noted above.

The solution obtained in this work is valid only for S-wave ($\ell = 0$). Nevertheless, by using the complex rotation (scaling) method [10], we obtain in Sec. 5 a highly accurate evaluation of the resonance structure associated with this potential for a given physical configuration and for *non-zero* angular momentum. In the same section, we utilize the J-matrix method to calculate the phase shift for the scattering of an electron from a molecule modeled by a given set of values of the potential parameters. But first, we start in the following section by implementing our approach in paper I to obtain an exact S-wave solution for the 3D problem with the potential (1.1). The reader is advised to consult Sec. I and Sec. II in paper I [4] for details on the theoretical formulation, background, and motivation of the approach. Section III in paper I gives an illustration on how to implement this approach to obtain an exact solution for the infinite potential well with sinusoidal bottom. A brief summary of the main findings in this paper was published recently in a letter [11].

## II. SOLUTION IN THE TRIDIAGONAL REPRESENTATION

The time-independent S-wave ($\ell = 0$) Schrödinger equation for a dynamical system modeled by a point particle of mass *m* in the field of a spherically symmetric potential *V(r)* is

$$\left[-\frac{\hbar^2}{2m}\frac{d^2}{dr^2} + V(r) - E\right]\psi(r, E) = 0, \qquad (2.1)$$

where $E$ is the particle's energy and $\psi(r,E)$ is the wavefunction that describes the state. Now, we make a transformation, $y = y(\lambda r)$, to a "reference" configuration space with coordinate $y \in [-1,+1]$, where $\lambda$ is a positive scale parameter having the dimension of inverse length. This transformation takes the wave equation (2.1) into

$$\left[ (y')^2 \frac{d^2}{dy^2} + y'' \frac{d}{dy} + U(y) \right] \psi(y,E) = 0, \tag{2.2}$$

where $U(y(\lambda r)) = \frac{2m}{\lambda^2 \hbar^2}[E - V(r)]$ and the prime stands for the derivative with respect to $\lambda r$. As shown in paper I, our choice of a complete square integrable basis that is compatible with this problem and carries a faithful description of the wavefunction, $\psi(y,E)$, has the following elements

$$\phi_n(y) = A_n (1+y)^\alpha (1-y)^\beta P_n^{(\mu,\nu)}(y), \tag{2.3}$$

where $P_n^{(\mu,\nu)}(y)$ is the Jacobi polynomial of degree $n = 0,1,2,..$ [12]. The parameters $\mu$ and $\nu$ are larger than $-1$ whereas the values of $\alpha$ and $\beta$ depend on the boundary conditions and square integrability. For bound states, they are real and mostly positive; however, for scattering states they may assume complex values. The normalization constant is chosen as $A_n = \sqrt{\frac{2n+\mu+\nu+1}{2^{\mu+\nu+1}} \frac{\Gamma(n+1)\Gamma(n+\mu+\nu+1)}{\Gamma(n+\nu+1)\Gamma(n+\mu+1)}}$. For a given physical system modeled by the potential function in $U(y)$, a crucial constraint on the choice of coordinate transformation, $y = y(\lambda r)$, is that the basis (2.3) should result in a tridiagonal matrix representation for the wave operator $J(y)$ in (2.2), where $J = (y')^2 \frac{d^2}{dy^2} + y'' \frac{d}{dy} + U(y)$. That is, the matrix elements $J_{nm} = \langle \phi_n | J | \phi_m \rangle$ must vanish for all $|n-m| \geq 2$. Stated differently, for a given coordinate transformation $y = y(\lambda r)$, the tridiagonal requirement limits the number of solvable potential functions in $U(y)$ to a very special set. We expand the wavefunction in the basis (2.3) as $\psi(y,E) = \sum_{n=0}^{\infty} f_n(E) \phi_n(y)$, where $\{f_n\}$ are the (Fourier) expansion coefficients. Therefore, a complete solution of the problem is obtained if all $\{f_n(E)\}_{n=0}^{\infty}$ are determined. As explained in paper I, it might be sufficient from the physical point of view (without going into rigorous mathematical analysis) to see that completeness and square integrability guarantee boundedness and convergence of this series. This will also be demonstrated numerically below. Now, if the tridiagonal constraint is satisfied then the wave equation (2.2) becomes equivalent to the following three-term recursion relation for the expansion coefficients

$$J_{n,n-1} f_{n-1} + J_{n,n} f_n + J_{n,n+1} f_{n+1} = 0. \tag{2.4}$$

With a proper choice of normalization, the solutions of this recursion relation are polynomials in some given physical parameter (e.g., the energy or potential parameter) [13]. For most cases in the class of solvable potentials obtained by our approach, the orthogonal polynomials associated with the resulting recursion relation do not belong to any of the known classic polynomials (e.g., the Hermite, Chebyshev, Laguerre, etc.) [14]. That is, their associated weight functions, generating functions, zeros, etc. are not found elsewhere. Nonetheless, they are completely defined for all degrees by their recursion relations and initial seed values. As stated in the Introduction, the theoretical formulation, background, and motivation for our "tridiagonal physics" approach could be found in paper I and in several other publications cited therein.

Next, we will show that the coordinate transformation $y = 1 - 2e^{-\lambda r}$ will meet the tridiagonal requirement for a suitable choice of potential function $U(y)$ and basis parameters. The integration measure is $\int_0^\infty ... dr = \frac{1}{\lambda} \int_{-1}^{+1} ... \frac{dy}{1-y}$ and the wave equation (2.2) becomes

$$\left[(1-y^2)\frac{d^2}{dy^2} - (1+y)\frac{d}{dy} + \frac{1+y}{1-y}U(y)\right]\psi(y,E) = 0, \tag{2.5}$$

where $\frac{1-y}{1+y}$ has been factored away. Since this factor vanishes at $y = 1$, then it might invalidate the wave equation as $r \to \infty$. Consequently, we must pay extra attention to the solution of (2.5) at the boundaries, which will be done below. The action of the wave operator on the basis elements (2.3) is calculated with the help of the differential equation, differential formula, and recursion relations of the Jacobi polynomials [12]. The result is as follows

$$\frac{1+y}{1-y}J|\phi_n\rangle = \left[-\alpha(2\beta+1) - n(n+\mu+\nu+1) - n\left(y + \frac{\nu-\mu}{2n+\mu+\nu}\right)\left(\frac{2\alpha-\nu-1}{1+y} + \frac{\mu-2\beta}{1-y}\right)\right.$$
$$\left. + \alpha(\alpha-1)\frac{1-y}{1+y} + \beta^2\frac{1+y}{1-y} + \frac{1+y}{1-y}U\right]|\phi_n\rangle + 2\frac{(n+\mu)(n+\nu)}{2n+\mu+\nu}\left(\frac{2\alpha-\nu-1}{1+y} + \frac{\mu-2\beta}{1-y}\right)\frac{A_n}{A_{n-1}}|\phi_{n-1}\rangle \tag{2.6}$$

The recursion relation and orthogonality formula of the Jacobi polynomials show that the matrix representation for the wave operator $\langle\phi_n|J|\phi_m\rangle$ becomes tridiagonal only in three cases:

$$\nu = 2\alpha - 1,\ \mu = 2\beta: \qquad \frac{1+y}{1-y}U = -A\frac{1-y}{1+y} - B\frac{1+y}{1-y} - Cy - D \tag{2.7a}$$

$$\nu = 2\alpha - 1,\ \mu = 2\beta - 1: \qquad \frac{1+y}{1-y}U = -A\frac{1-y}{1+y} - C\frac{y}{1-y} - \frac{D}{1-y} \tag{2.7b}$$

$$\nu = 2\alpha - 2,\ \mu = 2\beta: \qquad \frac{1+y}{1-y}U = -B\frac{1+y}{1-y} - C\frac{y}{1+y} - \frac{D}{1+y} \tag{2.7c}$$

where $A$, $B$, $C$, and $D$ are real dimensionless parameters and the basis parameters $\alpha$ and $\beta$ must satisfy $\alpha(\alpha-1) = A$ and $\beta^2 = B$. The second and third cases above correspond to the "generalized Hulthén" potential, which is the sum of the Hulthén potential and its square. This 3D S-wave problem is already known and its exact solution has been obtained and classified by many researchers [15]. Thus, we will not consider these two cases. On the other hand, the first case results in the following four-parameter potential function (in the units $\hbar = m = 1$)

$$\frac{2}{\lambda^2}V(r) = \frac{D}{e^{\lambda r} - 1} + \frac{A}{(e^{\lambda r} - 1)^2} + C\frac{1 - 2e^{-\lambda r}}{e^{\lambda r} - 1}, \tag{2.8}$$

And gives the basis parameter assignments: $\mu^2 = 4B = -8E/\lambda^2$ and $\nu^2 = 4A + 1$. Thus, possible real solutions of this problem are confined to negative energy (bound states) and for $A \geq -\frac{1}{4}$. Moreover, the basis (2.3) becomes energy dependent through the parameter $\mu(E)$ and the wave equation (2.5) becomes

$$\left[(1-y^2)\frac{d^2}{dy^2} - (1+y)\frac{d}{dy} - A\frac{1-y}{1+y} + \varepsilon\frac{1+y}{1-y} - Cy - D\right]\psi(y,\varepsilon) = 0, \tag{2.5'}$$

where $\varepsilon = 2E/\lambda^2$ is the energy parameter. Defining the dimensionless ratio $\gamma = \frac{D+C}{2C}$ for $C \neq 0$, we can rewrite the potential function (2.8) as

–5–

$$V(r) = \frac{\lambda^2 A/2}{(e^{\lambda r}-1)^2} + \lambda^2 C \frac{\gamma - e^{-\lambda r}}{e^{\lambda r}-1}. \tag{2.9}$$

Using the recursion relation and orthogonality formula of the Jacobi polynomials, we obtain the following matrix elements of the wave operator,

$$J_{nm} = \left[ D - A + \left(n + \tfrac{\nu+1}{2}\right)\left(n + \mu + \tfrac{\nu+1}{2}\right) \right]\delta_{nm} + C\langle n|y|m\rangle, \tag{2.10}$$

where $\langle n|y|m\rangle$ is the following tridiagonal matrix (see, the Appendix in paper I)

$$\begin{aligned}
\langle n|y|m\rangle &= \frac{\nu^2 - \mu^2}{(2n+\mu+\nu)(2n+\mu+\nu+2)}\delta_{n,m} \\
&+ \frac{2}{2n+\mu+\nu}\sqrt{\frac{n(n+\mu)(n+\nu)(n+\mu+\nu)}{(2n+\mu+\nu-1)(2n+\mu+\nu+1)}}\,\delta_{n,m+1} \\
&+ \frac{2}{2n+\mu+\nu+2}\sqrt{\frac{(n+1)(n+\mu+1)(n+\nu+1)(n+\mu+\nu+1)}{(2n+\mu+\nu+1)(2n+\mu+\nu+3)}}\,\delta_{n,m-1}
\end{aligned} \tag{2.11}$$

Therefore, the off-diagonal entries in the wave operator matrix (2.10) are due only to the last term, which is proportional to the potential parameter $C$. Hence, the diagonal representation constraint requires that $C = 0$ giving $\varepsilon_n = -\tfrac{1}{4}\left(n + \alpha + \tfrac{D-A}{n+\alpha}\right)^2$, where $\alpha(\alpha-1) = A$. This is the well-known energy spectrum formula for the generalized Hulthén potential [15]. However, we are interested in the new component of the potential (2.9) that corresponds to the special case where $A = 0$ and $C \neq 0$:

$$V(r) = -\lambda^2 C \frac{e^{-\lambda r} - \gamma}{e^{\lambda r}-1}. \tag{2.12}$$

Obviously, diagonalizing the Hamiltonian with this potential in the basis (2.3) will not lead to an exact solution because it requires $C = 0$. This is the reason why we relax the diagonal constraint by working in a more general tridiagonal representation that makes it possible to search for such a solution, if it existed. Now, for all finite values of the real parameter ratio $\gamma$, this potential has $1/r$ singularity at the origin and an exponentially decaying tail (i.e., it is short-range). Moreover, simple analysis of this potential function shows that the most interesting physical situation occurs when there is a local extremum (maximum or minimum) of the potential. That is, $dV/dr = 0$ at some finite radius. This situation is possible only if the value of $\gamma$ falls within the open range $0 < \gamma < 1$, which is equivalent to $|D| < |C|$. Thus, this potential becomes identical to the one given in the Introduction section by Eq. (1.1) with $V_0 = -\lambda^2 C$ and, thus, has all of its interesting features and physical interpretation. Additionally and as stated above, our approach can still handle the bound states solution for the case $\gamma \leq 0$ or $\gamma \geq 1$ provided that $\gamma V_0 > 0$. Now, taking $A = 0$ gives $\nu = +1$ and results in the following basis element

$$\phi_n(r) = \sqrt{\tfrac{n+\mu+1}{n+1}(2n+\mu+2)}\, e^{-\tfrac{1}{2}\lambda\mu r}\left(1-e^{-\lambda r}\right) P_n^{(\mu,1)}(1-2e^{-\lambda r}), \tag{2.13a}$$

where $\mu = 2\sqrt{-\varepsilon}$. Alternatively, we can write it in terms of the hypergeometric function ${}_2F_1\!\left({a,b \atop c}\big|z\right)$ as follows

$$\phi_n(r) = \tfrac{\Gamma(n+\mu+1)}{\Gamma(\mu+1)\Gamma(n+1)}\sqrt{\tfrac{n+\mu+1}{n+1}(2n+\mu+2)}\, e^{-\tfrac{1}{2}\lambda\mu r}\left(1-e^{-\lambda r}\right) {}_2F_1\!\left({-n, n+\mu+2 \atop \mu+1}\Big|e^{-\lambda r}\right), \text{ or} \tag{2.13b}$$

$$\phi_n(r) = (-1)^n \sqrt{(n+1)(n+\mu+1)(2n+\mu+2)}\, e^{-\tfrac{1}{2}\lambda\mu r}\left(1-e^{-\lambda r}\right) {}_2F_1\!\left({-n, n+\mu+2 \atop 2}\Big|1-e^{-\lambda r}\right). \tag{2.13c}$$

It is obvious that this function vanishes at the boundary ($r = 0$ and $r \to \infty$) for all energies, which settles our concern regarding the boundary conditions that was raised below the wave equation (2.5). One must remember, though, that an eigenfunction of the problem that corresponds to a given (negative) energy in the spectrum is an infinite sum of these elements that also depends on the potential parameters $C$ and $\gamma$ via the expansion coefficients $\{f_n\}$.

### III. POTENTIAL PARAMETER SPECTRUM VERSUS ENERGY SPECTRUM

Because the basis (2.13) is energy dependent, our solution strategy will differ from that in paper I. Here, we adopt the following scheme: For an arbitrarily chosen (negative) value of the energy, we find the set of values of the potential parameters that leads to an exact solution. Depending on the energy and physical constraints, this set could be finite or infinite. We call this set, the "potential parameter spectrum" or simply the *parameter spectrum*. The concept of a parameter spectrum was introduced for the first time in the solution of the wave equation in [3]. If the map that associates the parameter spectrum with the energy is invertible, then we could easily obtain the *energy spectrum* for a given choice of potential parameters. Now, for a fixed value of the energy $\varepsilon$ (equivalently, a constant non-negative parameter $\mu = 2\sqrt{-\varepsilon}$), the tridiagonal matrix representation of the wave operator (2.10) in the basis (2.13) becomes

$$J_{nm} = \left[ (2\gamma - 1)C + (n+1)(n+\mu+1) + \frac{(1-\mu^2)C}{(2n+\mu+1)(2n+\mu+3)} \right] \delta_{n,m}$$
$$+ \frac{2C}{2n+\mu+1} \sqrt{\frac{n(n+1)(n+\mu)(n+\mu+1)}{(2n+\mu)(2n+\mu+2)}} \delta_{n,m+1} + \frac{2C}{2n+\mu+3} \sqrt{\frac{(n+1)(n+2)(n+\mu+1)(n+\mu+2)}{(2n+\mu+2)(2n+\mu+4)}} \delta_{n,m-1}$$
(3.1)

Hence, the resulting three-term recursion relation for the expansion coefficients of the wavefunction becomes:

$$(1-2\gamma)f_n = \left(\tfrac{1}{C}a_n - d_n\right)f_n + b_{n-1}f_{n-1} + b_n f_{n+1}, \quad \text{where} \tag{3.2}$$

$$\left. \begin{array}{l} a_n(\varepsilon) = (n+1)(n+\mu+1), \quad d_n(\varepsilon) = \dfrac{\mu^2 - 1}{(2n+\mu+1)(2n+\mu+3)}, \\[6pt] b_n(\varepsilon) = \dfrac{2}{2n+\mu+3}\sqrt{\dfrac{(n+1)(n+2)(n+\mu+1)(n+\mu+2)}{(2n+\mu+2)(2n+\mu+4)}} \end{array} \right\}. \tag{3.3}$$

We write (3.2) as the eigenvalue equation $T_C |f\rangle = (1-2\gamma)|f\rangle$, where $T_C$ is the tridiagonal symmetric matrix $(T_C)_{nm} = \left(\tfrac{1}{C}a_n - d_n\right)\delta_{nm} + b_{n-1}\delta_{n,m+1} + b_n\delta_{n,m-1}$. Thus, for a given energy (equivalently, a given $\mu$) and potential strength $C$, this eigenvalue equation produces the $\gamma$-parameter spectrum. It is evident that the eigenvalue $(1-2\gamma)$, which for our class of problems is supposed to be limited to the open range $[-1,+1]$, grows rapidly with the size $N$ of the matrix $T_C$ as $N^2/C$ due to the entry $a_n$ in the diagonal term. The two limits $(1-2\gamma) \to \pm 1$ correspond to the two critical values $C^\pm$ of the potential strength parameter at which $D = \mp C$. In fact, there is an infinite number of these critical values at each energy. We write this set as $\{C_n^\pm(\varepsilon)\}$ where $\pm C_n^\pm(\varepsilon)$ are positive and each represents the minimum value of the potential strength for which an $n^{\text{th}}$ level bound state with energy greater than or equal to $-\varepsilon$ appears in the spectrum. To calculate these critical values for a given energy, we proceed as follows. We start by rewriting (3.2) as

$$-C^{-1} f_n = (a_n)^{-1} \left[ (2\gamma - 1 - d_n) f_n + b_{n-1} f_{n-1} + b_n f_{n+1} \right], \tag{3.4}$$

If we define the new coefficients $g_n^\mu = \sqrt{a_n/a_0}\, f_n$, then we can write this recursion relation in terms of them as

$$-C^{-1} g_n^\mu = \mathcal{A}_n\, g_n^\mu + \mathcal{B}_{n-1}\, g_{n-1}^\mu + \mathcal{B}_n\, g_{n+1}^\mu, \quad n = 1, 2, \ldots, \tag{3.5}$$

where $\mathcal{A}_n = (2\gamma - 1 - d_n)/a_n$ and $\mathcal{B}_n = b_n/\sqrt{a_n a_{n+1}}$. The initial relation ($n = 0$) for this recursion is

$$-C^{-1} g_0^\mu = \mathcal{A}_0\, g_0^\mu + \mathcal{B}_0\, g_1^\mu. \tag{3.6}$$

The new recursion coefficients $\mathcal{A}_n$ and $\mathcal{B}_n$ approach the limit of large $n$ as $n^{-2}$. Thus, using (3.5) to calculate the *C*-parameter spectrum gives a more rapidly convergent result than using (3.2) to calculate the $\gamma$-parameter spectrum. Figure 3 shows the lowest critical potential strength $\{C_n^\pm(\varepsilon)\}_{n=0}^{n=4}$ for all bound states in a given energy range. We were able to obtain these values, as solutions of Eq. (3.5) when written as the eigenvalue equation, $T_\gamma |g^\mu\rangle = -C^{-1}|g^\mu\rangle$ with $\gamma = 0$ or 1, to machine precision[†] with as low matrix size as 20×20. Out of these critical values, the most important for the system are those at zero energy (i.e., at the boundary of the energy spectrum). We refer to this subset by the symbol $\{\hat{C}_n\}$. In Table 1, we list some of these values displayed conveniently to an accuracy of 10 decimal places. In fact, for each $\gamma$ one can calculate these critical values of the potential strength at which a bound state gets created or destroyed. At these critical values, the state experiences a transition from bound to resonance or vice versa. Recently, this phenomenon was also demonstrated for the Yukawa potential and the transition process was displayed as video animation showing the trajectories of the energy eigenvalues in the complex energy plane [16]. Table 2, gives a list of the lowest positive and negative values in the set $\{\hat{C}_n(\gamma)\}$ for several choices of $\gamma$. An obvious relation between the two sets of critical potential strength is: $\hat{C}_n(0) = C_n^+(0)$ and $\hat{C}_n(1) = C_n^-(0)$. It will be evident from the following analysis of the energy spectrum that for a given parameter $\gamma$, a potential strength $C$ that lies in the range $\hat{C}_n(\gamma) < C < \hat{C}_{n+1}(\gamma)$ will result in $n+1$ bound states. Therefore, these critical values are very important for *bound states number counting*.

Now, since $\lambda$ is the only dimensionful parameter in the problem, then it might be obvious that the only possible dependence of the energy spectrum on $\lambda$ is via an overall factor of $\lambda^2$. Another way to see that, is by writing the radial Schrödinger equation (2.1) with the potential (2.12) and then rescaling the radial coordinate and energy as $r \to r/\lambda$ and $E \to \lambda^2 E$ causing the parameter $\lambda$ to disappear from the equation. For this reason, we work with the energy variable $\varepsilon = 2E/\lambda^2$ rather than $E$ so that $\lambda$ disappears and we deal only with two parameters, $C$ and $\gamma$, instead of three. To obtain the potential strength *parameter spectrum* for fixed values of $\gamma$ and for all energies in a conveniently chosen range, we solve the recursion relation (3.5) as the eigenvalue equation $T_\gamma |g^\mu\rangle = -C^{-1}|g^\mu\rangle$. Figure 4 shows the result of this calculation for a given choice of $\gamma$ and for energy ranges that are chosen appropriately depending on the sign of the resulting potential strength $V_0$. The figure is shown with $C$ on the horizontal axis and $-\varepsilon$ on the vertical axis to make it more convenient to visualize the energy spectrum. Thus, a vertical line that crosses the $C$-axis at a value, say, $\bar{C}$ intersects the curves at the energy spectrum corresponding to the problem with potential parameters $\bar{C}$ and $\gamma$. The figure shows that for a given $\gamma$, the spacing of the energy spectrum is larger for positive $C$ than for negative $C$ (i.e., the energy spectrum is denser for negative $C$ but is more stretched for positive $C$). In fact, this should have already been obvious from the physics of the potential (2.12), as portrayed in Fig. 1a and Fig. 1b. Moreover, reproducing Fig. 4 for different values of $\gamma$ leads to the following conclusion about another interesting property of the energy spectrum: For a given $|C|$, the energy spectrum is larger for positive $C$ when $\gamma < \frac{1}{2}$ but larger for negative $C$ when $\gamma > \frac{1}{2}$.

---

[†] 15 significant digits on our laptop using Mathcad® 2000 computational software.

For a fixed $\gamma$, we can make a fit on each trace in the energy versus $C$-parameter spectrum of Fig. 4 with an $M^{\text{th}}$ order continued fraction using the rational fraction approximation of Haymaker and Schlessinger similar to that in the Padé method [17]. Consequently, the bound state energy could be written as a function of $C$ and $n$ for that fixed $\gamma$. In other words, given the potential parameters $C$ and $\gamma$, we can use this continued fraction to obtain the energy spectrum $\{\varepsilon_n(\gamma, C)\}$. Table 3 illustrates the computational stability and convergence of the energy spectrum with the order of the continued fraction. The Mathcad program codes used for this and other calculations in this work are available upon request from the author.

Finally, it is worth noting that the same process above could be repeated to obtain the *γ-parameter spectrum* using the recursion relation (3.2) for a fixed potential strength $C$ and a chosen energy range. Figure 5 shows such results for unconstrained range of $\gamma$.

### IV. BOUND STATES WAVEFUNCTION

The solution of the three-term recursion relation (3.5) for a given energy is defined modulo an overall non-singular function of the potential parameters $C$ and $\gamma$. If we call this function $\omega^\mu(\gamma, C)$, then we can write $g_n^\mu(C) = \omega^\mu(\gamma, C) Q_n^\mu(C)$. Substituting this in the recursion (3.5) with its initial relation (3.6) and choosing the standard normalization, $Q_0^\mu = 1$, determines $Q_n^\mu(C)$ as polynomials of degree $n$ in $C^{-1}$ for all $n$. For example, the first few are

$$Q_0^\mu(C) = 1 \tag{4.1a}$$

$$Q_1^\mu(C) = \tfrac{-1}{\mathcal{B}_0}\left(C^{-1} + \mathcal{A}_0\right) \tag{4.1b}$$

$$Q_2^\mu(C) = \tfrac{1}{\mathcal{B}_0 \mathcal{B}_1}\left[\left(C^{-1} + \mathcal{A}_0\right)\left(C^{-1} + \mathcal{A}_1\right) - \mathcal{B}_0^2\right] \tag{4.1c}$$

………..
………..

$$Q_n^\mu(C) = \tfrac{-1}{\mathcal{B}_{n-1}}\left[\left(C^{-1} + \mathcal{A}_{n-1}\right) Q_{n-1}^\mu(C) + \mathcal{B}_{n-2} Q_{n-2}^\mu(C)\right] \tag{4.1d}$$

We show below that completeness of the basis and normalization of the wavefunction give $\omega^\mu(\gamma, C) = 1/\sqrt{\mathcal{K}^\mu(\gamma)}$, where $\mathcal{K}^\mu(\gamma)$ is the kernel operator associated with these polynomials at the infinite order limit. Now, the polynomial $Q_n^\mu(C)$ does not belong to any of the known classes of orthogonal polynomials. However, if we define the polynomial $P_n^\mu(\gamma) = \sqrt{a_0/a_n}\, Q_n^\mu(C)$, then we can write $f_n(\varepsilon) = \omega^\mu(\gamma,C) P_n^\mu(\gamma)$ and $P_n^\mu(\gamma)$ satisfies the recursion relation (3.2) with the initial seed value $P_0^\mu(\gamma) = 1$. It is a polynomial in $(1-2\gamma)$ and satisfies the same recursion relation typical of those that we have found recently while attempting to find an extended class of solutions to the generalized Hulthén and Rosen-Morse problems [14] (see, Eq. (6.9) and Eq. (7.3)).

The theory of orthogonal polynomials gives an alternative means for obtaining the parameter spectrum [13]. It goes as follows: The *C*-parameter (or *γ*-parameter) spectrum is obtained form the zeros of the polynomial $Q_n^\mu(C)$ $\left(\text{or } P_n^\mu(\gamma)\right)$ at a given $\mu$ in the limit as $n \to \infty$, respectively. In fact, the calculated values of the lowest part of the spectrum converge quickly for relatively low polynomial degrees (as low as $n = 20$ with 12 digits accuracy).

Figure 6 is a plot of the lowest bound state energy eigenfunctions [normalized by $\omega^\mu(\gamma,C)$] for a given set of potential parameters *C* and *γ*. A wavefunction that corresponds to a bound state with energy $\varepsilon_n = 2E_n/\lambda^2$ is computed as $\psi(r,\varepsilon_n) \approx \omega^{\mu_n}(\gamma,C) \sum_{m=0}^{N-1} P_m^{\mu_n}(\gamma)\, \phi_m(r)$, for some large enough integer *N* and where *C* and *γ* belong to the parameter spectrum associated with $\varepsilon_n$. Numerically, we find that the sum converges quickly but becomes unstable if the number of terms, *N*, becomes too large exceeding an integer that depends on the potential parameters and energy level. For the choice of parameters in Fig. 6, our numerical routine produced the stable plots shown for $\psi(r,\varepsilon_n)/\omega^{\mu_n}(\gamma,C)$; but as *N* is increases beyond $N = 15$ it becomes unstable. Better numerical routines with higher degree of precision might be developed such that instability occurs at larger values of *N*. Moreover, trying to evaluate the wavefunction at an energy that does not belong to the energy spectrum will never achieve stable results. It will only produce rapidly increasing oscillations with large amplitudes. In fact, the sum of these oscillations for large *N* leads to destructive interference that should result in zero net value for the wavefunction. Now, the norm of the wavefunction at an energy eigenvalue is evaluated as follows

$$1 = \|\psi_k\| = \lambda \int_0^\infty |\psi(r,\varepsilon_k)|^2 dr = [\omega^{\mu_k}(\gamma,C)]^2 \sum_{n,m=0}^\infty \Theta_{nm} P_n^{\mu_k}(\gamma) P_m^{\mu_k}(\gamma), \qquad (4.2)$$

where $\Theta_{nm}$ is the basis overlap matrix $\langle \phi_n | \phi_m \rangle = \langle n | \frac{1+y}{1-y} | m \rangle$ [18]. Aside from $\Theta_{nm}$, which is symmetric and generally not separable in the indices as $\theta_n \theta_m$, this sum is the usual kernel for orthogonal polynomials [13]. That is, for the polynomials $p_n(z)$ associated with orthogonal representation, this is usually written as $K_N(z,z') = \sum_{n=0}^{N-1} p_n(z) p_n(z')$. Therefore, for orthogonal representations, where $\Theta_{nm} \sim \delta_{nm}$, the sum in (4.2) becomes the infinite limit of the usual kernel, $K(z) = \lim_{N\to\infty} K_N(z,z)$. However, for non-orthogonal representations, the kernel is evaluate as $\mathcal{K}_N(z,z') = \sum_{n,m=0}^{N-1} \Theta_{nm}\, p_n(z) p_m(z')$. Thus, Eq. (4.2) gives $\omega^\mu(\gamma,C) = 1/\sqrt{\mathcal{K}^\mu(\gamma)}$ and we write the bound state wavefunction as

$$\psi(r,\varepsilon_n) = \sqrt{\frac{2(\mu_n+1)}{\mathcal{K}^{\mu_n}(\gamma)}}(1-e^{-\lambda r})e^{-\frac{1}{2}\lambda\mu_n r}\sum_{m=0}^{\infty}\frac{1}{m+1}\sqrt{m+1+\frac{1}{2}\mu_n}\ Q_m^{\mu_n}(C)P_m^{(\mu_n,1)}(1-2e^{-\lambda r})\,, \quad (4.3)$$

where $\mu_n = 2\sqrt{-\varepsilon_n}$.

## V. ENERGY RESONANCE AND SCATTERING PHASE SHIFT

The solution of the 3D problem with the radial potential (1.1) that we have obtained using our tridiagonal representation approach is valid only for S-wave ($\ell = 0$) bound sates. In this section and for non-zero angular momentum, we obtain a highly accurate evaluation of the bound state and resonance energies for this 1/r singular potential using the complex rotation method in the tridiagonal J-matrix basis [19]. Recently, we have applied this method successfully in obtaining the bound state and resonance structure for the Yukawa potential [16]. The details of implementation of the method on 1/r singular short-range potentials are found in [19]. Figure 7 shows the spectrum of the potential in the complex energy plane for several choices of angular momentum using a finite dimensional Hamiltonian representation. The spectrum consists of:
  (i) Discrete points on the negative real axis that correspond to the bound states;
  (ii) A clockwise-rotated *line* of dense points approximating the rotated branch cut (discontinuity) of the finite Green's function; and
  (iii) Exposed resonances shown as well-separated points in the lower half of the complex energy plane located in the sector bound by the rotated cut line and the positive energy axis.

The location of points corresponding to bound states and resonances remain stable against variations in all nonphysical computational parameters [20]. Table 4 lists the bound states and resonance energies for several physical configurations (different values of $C$, $\gamma$, and $\ell$). On the other hand, Table 5 compares the bound states energy spectrum for $\ell = 0$ obtained independently by the finite complex rotation method and the parameter spectrum method given in Table 3.

Next, we obtain the phase shift for electron scattering off an extended molecule modeled by the potential (1.1) using the J-matrix method [6]. This method is an algebraic method of quantum scattering that gives exact scattering information for a model potential represented by its matrix elements in a finite subset of a complete square integrable basis. The basis is chosen such that it supports an infinite tridiagonal matrix representation of the reference Hamiltonian $H_0$, which is the part of the total Hamiltonian that is exactly solvable. Therefore, $H_0$ is accounted for analytically exactly and in full whereas the contribution of the potential is approximated by its finite matrix representation. Now, the potential function (1.1) is 1/r singular, where $\lim_{r\to 0}V(r) = Z_{eff}/r$. Then its matrix elements, which are obtained as integrals over the $L^2$ basis, may have large errors. To deal with this problem, we absorb the 1/r singularity of the potential into $H_0$, which can still be handled exactly in the J-matrix method. Therefore, we rewrite the total Hamiltonian as $H = \tilde{H}_0 + \tilde{V}$, where $\tilde{H}_0 = H_0 + Z_{eff}/r$ and $\tilde{V} = V - Z_{eff}/r$. Then $\tilde{V}$ becomes regular everywhere resulting in accurate evaluation of its matrix elements. The details of this scheme is given in [19]. Figure 8 shows the scattering phase shift for a chosen physical configuration. The figure indicates strong resonance activity around $\varepsilon = 4.0$. Detailed investigation confirms the presence of a sharp resonance at $\varepsilon = 4.03492 - i\,0.01465$.

## VI. CONCLUSION

In this article, which is the second in the series, we have succeeded in enlarging the class of solvable problems in 3D by adding to it the radial potential (1.1), which is 1/r singular and short-ranged. We achieved that by working in a complete square integrable basis that supports a tridiagonal matrix representation for the wave operator. This makes the wave equation equivalent to a three-term recursion relation for the expansion coefficients of the wavefunction. Consequently, finding a solution of the recursion relation is equivalent to solving the original problem. This method gives a larger class of solvable potentials. The usual diagonal representation constraint results in a reduction to the conventional class of solvable potentials. We found that the radial potential (1.1) meets the tridiagonal requirement for S-wave problems. Moreover, using the complex rotation method in a finite basis, we were also able to obtain a highly accurate evaluation of the resonance and bound states structure associated with this potential for several non-zero angular momenta. The physical properties of the potential and the structure of its spectrum are highly non-trivial. We argued that it could be very useful as a more appropriate model for the interaction of an electron with extended molecules whose electron cloud is congregated near the center of the molecule. As an illustration, we used the J-matrix method to calculate the phase shift for the scattering of an electron from a molecule modeled by a given set of values of the potential parameters. In this work, we also reintroduced the concept of a *parameter spectrum* where an exact solution of the problem is obtained at a single energy but for an infinite set of values of the potential parameters (the parameter spectrum). We found that the map that associates the parameter spectrum with the energy is invertible, thus we were able to obtain the *energy spectrum* for a given choice of potential parameters. We also defined the notion of *exact solvability* to be the ability to write the wavefunction in a closed form as a convergent series in terms of orthogonal polynomials. These polynomials, which are functions of the configuration space and of the energy, are well-defined to all orders. Additionally, all physical quantities in the problem (e.g., the energy spectrum, phase shift, wavefunction, resonances, etc.) are obtained to any desired accuracy limited only by the computing machine precision; no physical approximations are invoked.

In the near future, we will also report on an exact solution for the one-dimensional single-wave potential $V(x) = V_0 \left[ \tanh(\lambda x) + \gamma \right] / \cosh^2(\lambda x)$ with $-1 < \gamma < 1$, which has no previously known exact solution. Finally, it is worth noting that the formulation of this approach could easily be extended to noncentral [21] as well as relativistic problems. In such relativistic extension, one searches for a tridiagonal matrix representation of the Dirac operator in a suitable spinor basis similar to what has been done in [22].

## ACKNOWLEDGEMENTS

This work is sponsored by the Saudi Center for Theoretical Physics. Partial support by King Fahd University of Petroleum and Minerals under project SB-090001 is highly appreciated.

**Table captions:**

**Table 1**: Lowest set of the critical potential strength at zero energy, $C_n^\pm(0)$, to an accuracy of 10 decimal places.

**Table 2**: The smallest set of values of the critical potential strength $\hat{C}_n(\gamma)$ for several choices of $\gamma$.

**Table 3**: Convergence of the values of the lowest part of the spectrum (for a given $C$ and $\gamma$) with the continued fraction fit order $M$.

**Table 4:** Bound states and resonance energies associated with the potential (1.1) for several values of $C$, $\gamma$, and angular momentum $\ell$. These values were obtained by the complex scaling (rotation) method.

**Table 5:** Compares the bound states energy spectrum for $\ell = 0$ obtained independently by the finite complex scaling method and the parameter spectrum method of Table 3.

**Figure captions:**

**Fig. 1:** The potential function (1.1) in units of $|V_0|$ versus the radial coordinate in units of $\lambda^{-1}$: (a) for $V_0 > 0$ and for several values of $\gamma$ within the range $0 < \gamma < 1$, (b) for $V_0 < 0$ and $\gamma = 1/2$, and in (c) we show the effective potential, which is the sum of $V(r)$ for $V_0 < 0$ and the orbital term with non-zero angular momentum ($\ell = 3$ and $\gamma = 1/2$).

**Fig. 2:** Electron interacting with an extended molecule whose electron cloud is congregated near the center of the molecule (for example, due to a large centered atom).

**Fig. 3:** The lowest critical potential strength parameters $\{C_n^\pm(\varepsilon)\}_{n=0}^{n=4}$ for bound states energies in the range $0 \geq \varepsilon \geq -20$.

**Fig. 4:** The energy spectrum as a function of the potential strength parameter $C$ (positive and negative) for $\gamma = 0.2$.

**Fig. 5:** The energy spectrum as a function of unconstrained values of $\gamma$ for $C = 200$. Note that the solvability condition for values of $\gamma$ outside the range $0 < \gamma < 1$, which is $\gamma V_0 > 0$, is satisfied.

**Fig. 6:** A plot of the lowest four energy eigenfunctions $\psi(r, \varepsilon_n)$ [normalized by $\omega^{\mu_n}(\gamma, C)$] versus the radial coordinate (in units of $\lambda^{-1}$) for $\gamma = 0.7$, $C = -70$.

**Fig. 7:** The potential spectrum (bound states and resonance energies) in the complex $\varepsilon$-plane for $\gamma = 0.5$, $C = 80$, and for several values of the angular momentum. Bound states (Resonances) are shown as boxed (circled) dots, while the string of bare dots represents the rotated cut *line* (discontinuity of the Green's function). One (two) bound state energy (energies) is (are) outside the range of the figure with $\ell = 1$ ($\ell = 0$) on the negative real line.

**Fig. 8:** P-wave single-channel scattering phase shift for an electron off an extended molecule modeled by the parameters $Z = V_0/\lambda = -70$ and $Z_{eff} = V_0(1-\gamma)/\lambda = -42$ and $\lambda = 1.0$ a.u. (i.e., $\gamma = 0.4$ and $C = 70$). The presence of sharp resonance around $\varepsilon = 4.0$ is very clear.

**Table 1**

| $n$ | $C_n^+(0)$ | $C_n^-(0)$ |
|---|---|---|
| 0 | 1.2956609331 | −0.7228982454 |
| 1 | 5.0184325653 | −3.8089077930 |
| 2 | 11.1997215264 | −9.3608758488 |
| 3 | 19.8446859831 | −17.3800355533 |
| 4 | 30.9550078158 | −27.8665379522 |
| 5 | 44.5314400641 | −40.8204191165 |
| 6 | 60.5743842474 | −56.2416910648 |
| 7 | 79.0840796714 | −74.1303587070 |
| 8 | 100.0606804461 | −94.4864243480 |
| 9 | 123.5042916444 | −117.3098891845 |
| 10 | 149.4149881179 | −142.6007538875 |

**Table 2**

| | $n$ | $\hat{C}_n(0.2)$ | $\hat{C}_n(0.4)$ | $\hat{C}_n(0.6)$ | $\hat{C}_n(0.8)$ |
|---|---|---|---|---|---|
| | 0 | 2.2152611940 | 4.4806954308 | 11.0749939486 | 47.3560824553 |
| | 1 | 9.1241352235 | 18.2554807127 | 44.4781677038 | 189.5457747849 |
| $V_0 < 0$ | 2 | 20.7863387221 | 41.1954191471 | 100.1462183676 | 426.5282857052 |
| | 3 | 37.1341458455 | 73.3066806258 | 178.0810805327 | 758.3037612186 |
| | 4 | 58.1482610684 | 114.5915546473 | 278.2829383518 | 1184.8722188768 |
| | 5 | 83.8267659355 | 165.0504933315 | 400.7518350936 | 1706.2336629858 |
| | 0 | −12.5836736341 | −3.2560380676 | −1.5789834905 | −0.9994647235 |
| | 1 | −107.3824443481 | −25.5165055798 | −10.7688457640 | −5.8706836142 |
| $V_0 > 0$ | 2 | −296.9689222484 | −70.0502840866 | −29.1082728662 | −15.2315548361 |
| | 3 | −581.3480285245 | −136.8514039954 | −56.6289478043 | −29.2235397247 |
| | 4 | −960.5200426162 | −225.9196386161 | −93.3253687332 | −47.8823486902 |
| | 5 | −1434.4850177533 | −337.2549550286 | −139.1965109280 | −71.2132815834 |

**Table 3**

| $n$ | $M = 10$ | $M = 20$ | $M = 50$ | $M = 100$ |
|---|---|---|---|---|
| 0 | 70.014054905396 | 70.014054905332 | 70.014054905332 | 70.014054905331 |
| 1 | 50.181498523641 | 50.181498523547 | 50.181498523546 | 50.181498523546 |
| 2 | 34.317359872296 | 34.317359873422 | 34.317359873422 | 34.317359873422 |
| 3 | 21.924290023027 | 21.924290020805 | 21.924290020805 | 21.924290020806 |
| 4 | 12.606338995716 | 12.606339023394 | 12.606339023389 | 12.606339023389 |
| 5 | 6.041070541005 | 6.041070158135 | 6.041070158115 | 6.041070158115 |
| 6 | 1.960936292059 | 1.960935896912 | 1.960935939333 | 1.960935939299 |
| 7 | 0.140393239779 | 0.140375436793 | 0.140389006550 | 0.140389009571 |

**Table 4**

| $\ell$ | $\gamma = 0.3, C = 50$ | $\gamma = 0.5, C = 80$ | $\gamma = 0.7, C = 100$ |
|---|---|---|---|
| 0 | | −1406.11040577 | −679.95986643 |
| | | −223.29635015 | −32.96147955 |
| | −1094.42109160 | −27.18320883 | 44.53768627 −i 4.67586216 |
| | −187.97359168 | 14.78518500 −i 1.61589438 | 43.54320167 −i 28.99664646 |
| | −36.02622806 | 14.74796320 −i 14.45111009 | 30.49825604 −i 49.95491445 |
| | 2.49135906 −i 4.44858137 | 5.87666008 −i 26.32754872 | 12.03861742 −i 65.17158354 |
| | −3.47269075 −i 9.76206122 | −7.73059099 −i 34.45981639 | −9.22487684 −i 75.50096443 |
| 1 | | −219.66959141 | −27.29186980 |
| | −185.38841241 | −24.21918006 | 48.52849166 −i 5.44855836 |
| | −33.95322592 | 16.59977495 −i 2.02268673 | 46.97743366 −i 30.97381334 |
| | 1.90006620 −i 0.05866355 | 16.14919395 −i 15.63387859 | 33.41744568 −i 52.74946446 |
| | 3.10281186 −i 5.30008439 | 6.89969999 −i 27.98583841 | 14.45887019 −i 68.66780260 |
| | −3.13152673 −i 10.95482666 | −7.07427984 −i 36.52772943 | −7.29348425 −i 79.62908216 |
| 2 | | | 58.38580965 −i 6.72074273 |
| | | −18.01714564 | 56.28521956 −i 34.19445283 |
| | −29.64332195 | 20.63831671 −i 2.95009034 | 42.35875422 −i 57.41783863 |
| | 4.38214060 −i 0.32914421 | 19.47023201 −i 18.15649888 | 23.16581798 −i 74.61291728 |
| | 4.67860987 −i 7.15177747 | 9.56170137 −i 31.62847960 | 1.28209365 −i 86.72210727 |
| | −2.17410775 −i 13.60827617 | −5.03946029 −i 41.20500393 | −22.13656236 −i 94.64163955 |

**Table 5**

| $n$ | Parameter Spectrum | Complex Scaling |
|---|---|---|
| 0 | 70.014054905331 | 70.014054905331 |
| 1 | 50.181498523546 | 50.181498523549 |
| 2 | 34.317359873422 | 34.317359873422 |
| 3 | 21.924290020806 | 21.924290020806 |
| 4 | 12.606339023389 | 12.606339023389 |
| 5 | 6.041070158115 | 6.041070158115 |
| 6 | 1.960935939299 | 1.960935939298 |
| 7 | 0.140389009571 | 0.140389009245 |

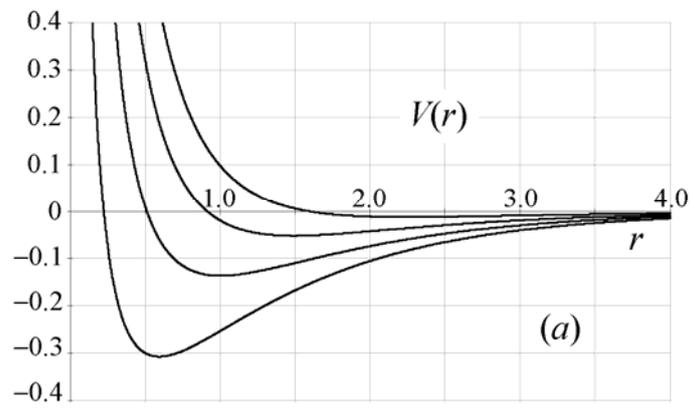

**Fig. 1a**

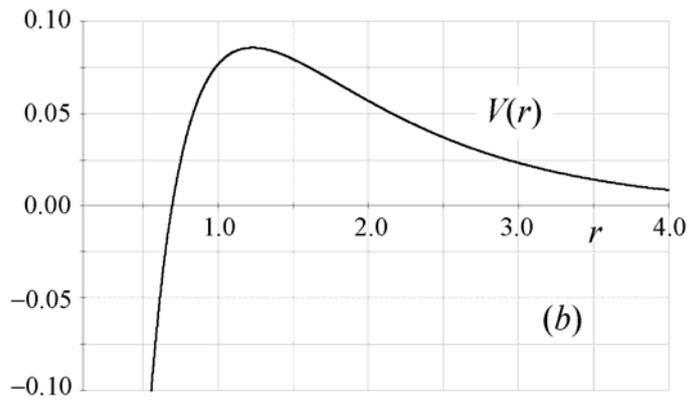

**Fig. 1b**

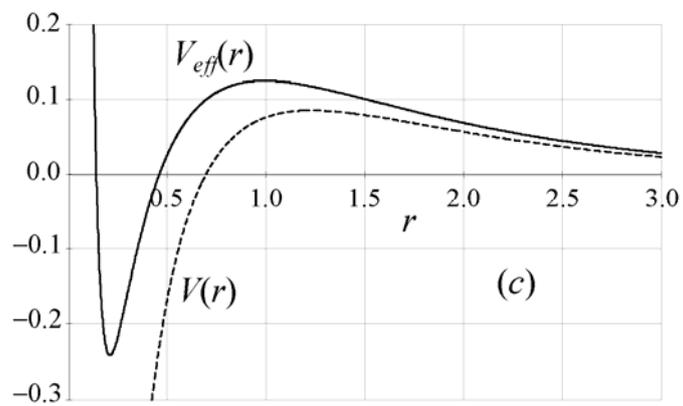

**Fig. 1c**

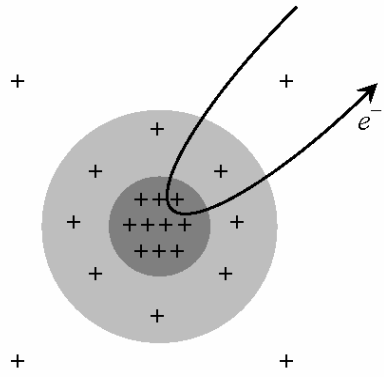

**Fig. 2**

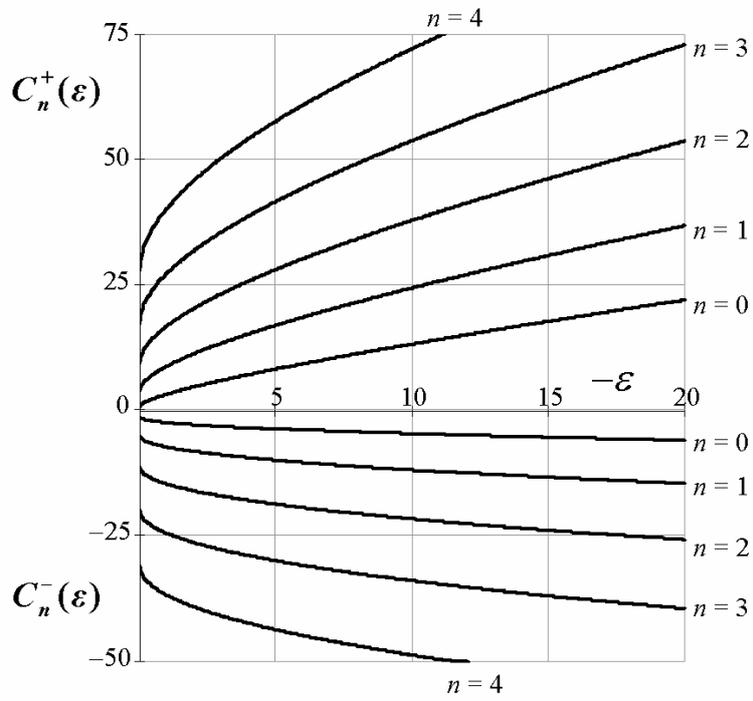

**Fig. 3**

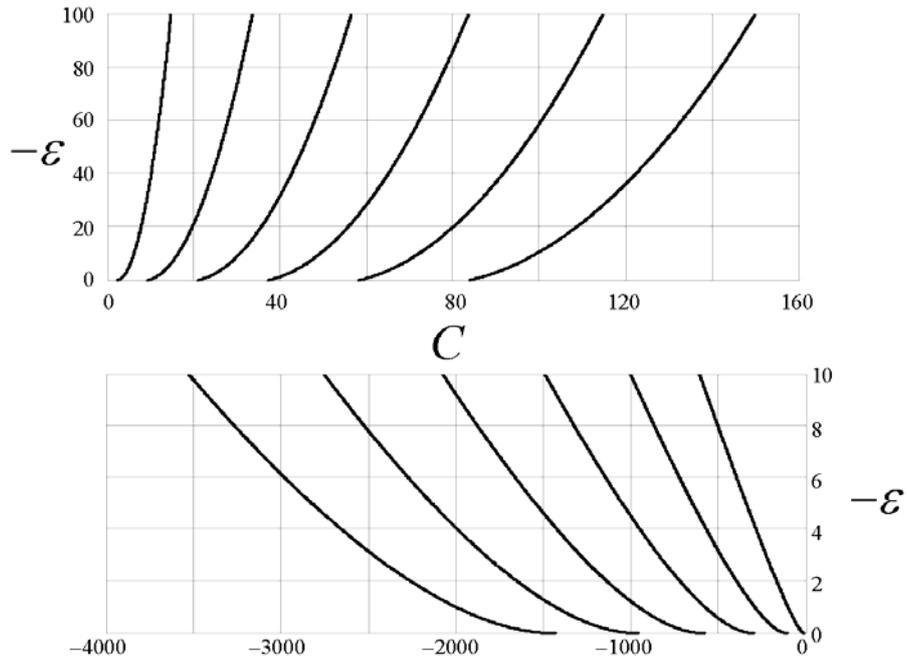

**Fig. 4**

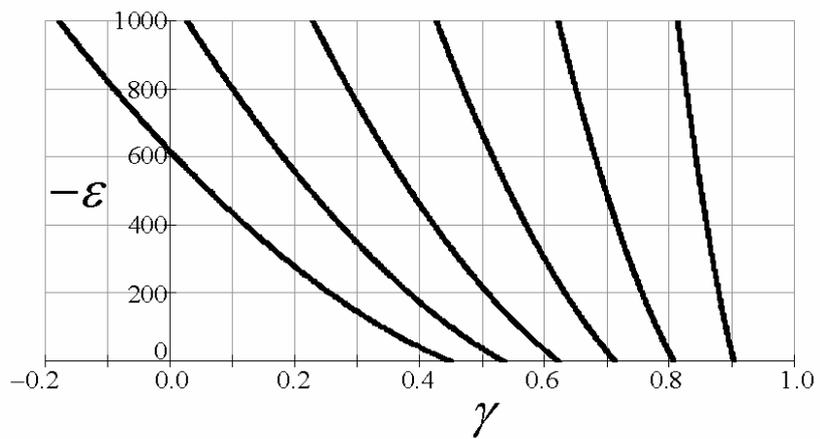

**Fig. 5**

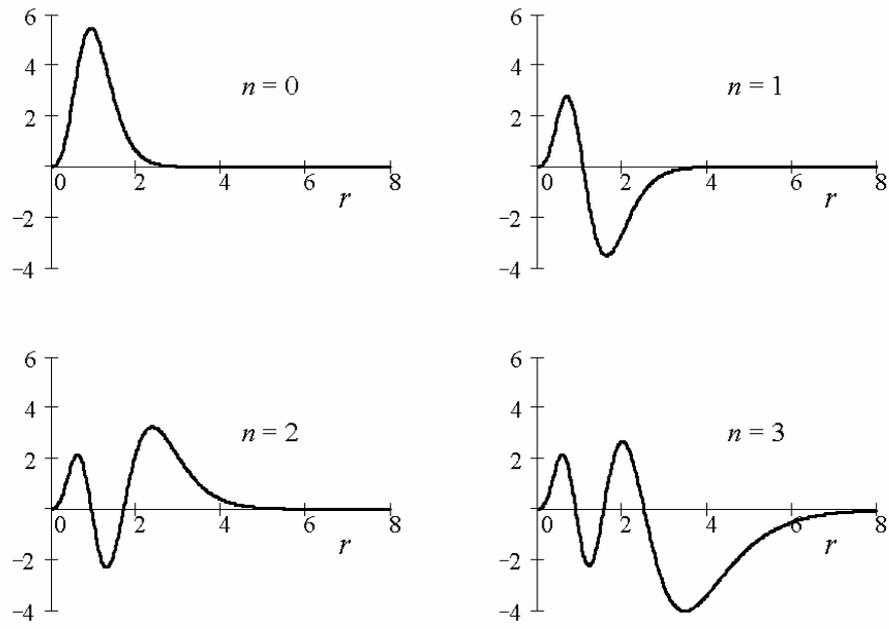

**Fig. 6**

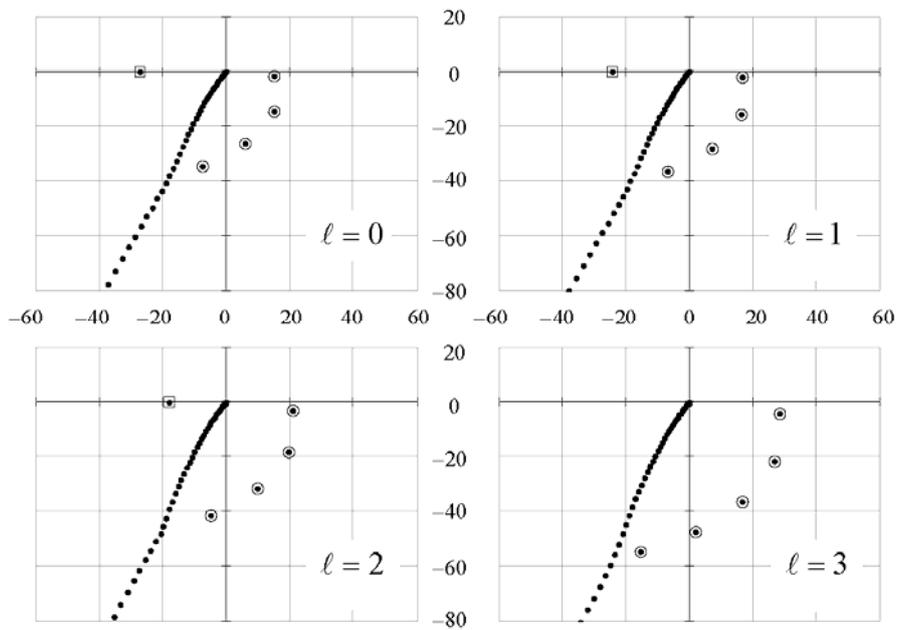

**Fig. 7**

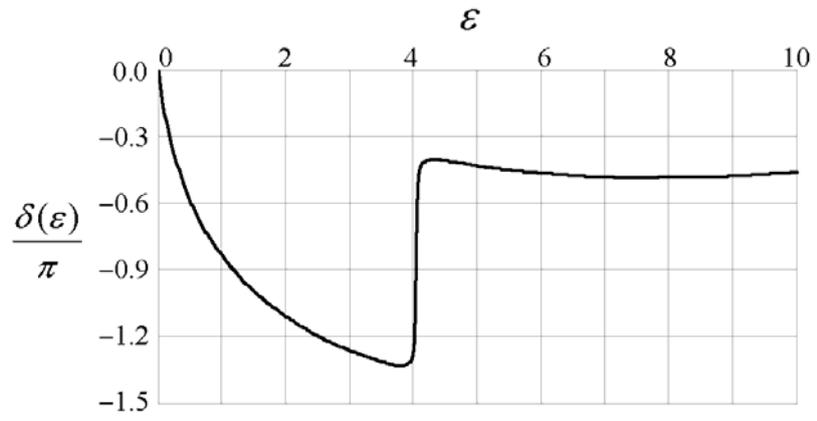

**Fig. 8**